\documentclass{article}
\usepackage{spconf,amsmath,graphicx,hyperref}
\usepackage{amssymb}

\usepackage{booktabs}

\title{Beyond Textual Chain-of-Thought: JEPA-Conditioned Latent Reasoning for Large Audio Language Models}
\name{%
\begin{tabular}{@{}c@{}}
\textit{Donghang Wu$^{1,2}$, Haoyang Zhang$^{3}$, Yizhou Peng$^{1}$, Shreyas Gopal$^{1}$, 
 Yi-Wen Chao$^{1}$,}\\
\textit{Chen Chen$^{1}$, Hexin Liu$^{1}$, William Tjhi$^{1,2}$, Eng-Siong Chng$^{1,*}$\thanks{*Corresponding author: \texttt{aseschng@ntu.edu.sg}}}
\end{tabular}%
}

\address{%
$^{1}$ Nanyang Technological University, Singapore\\
$^{2}$ AI Singapore, Singapore\\
$^{3}$ Peking University, Beijing, China
}
\begin{document}
%
\maketitle
\begin{abstract}
Explicit textual Chain-of-Thought (CoT) has improved the reasoning ability of large audio language models (LALMs). However, textual CoTs are often constructed from text captions of audio and provide limited access to the acoustic evidence, which can introduce problems like hallucination. To address this modality-gap issue, we introduce JELAR, a Joint-Embedding Predictive Architecture (JEPA)-based latent reasoning framework that conditions latent reasoning supervision on acoustic representations learned from raw waveforms. During training, a frozen WavJEPA model provides representations learned directly from raw waveforms. A non-causal expert first constructs answer-aware queries, which cross-attend to WavJEPA embeddings to produce latent reasoning targets. The LALM is trained to predict these targets before generating its response. Experimental results show that JELAR improves the Audio-Reasoner baseline by 2.70 and 9.10 absolute percentage points on MMAU-mini and MMAR, respectively, demonstrating the effectiveness of JEPA-conditioned latent reasoning as an alternative to explicit textual CoT supervision.
\end{abstract}
\begin{keywords}
Large Audio Language Model, Latent Reasoning, Chain-of-Thought, WavJEPA,
Audio Reasoning
\end{keywords}
\section{Introduction}
\label{sec:intro}

Large audio language models (LALMs) have recently demonstrated increasingly strong capabilities in speech understanding, spoken dialogue, environmental sound analysis, music perception, and audio question answering \cite{qwen2audio, team2026qwen3, chronological}. Nevertheless, reasoning over audio remains substantially more challenging than recognizing isolated acoustic events \cite{mps, stepaudior1}. Complex tasks often require a model to identify relevant evidence, integrate information across time, relate sound sources, and combine acoustic perception with semantic inference \cite{audioreasoningsurvey, audiothinker, audioreasoner}.

Explicit chain-of-thought (CoT) has emerged as a promising approach to improving such capabilities \cite{audiothinker, audioreasoner}. Audio-Reasoner \cite{audioreasoner}, for example, constructs structured textual reasoning traces and trains an LALM to generate a CoT before its final response. However, recent studies have highlighted a fundamental modality-gap concern in audio reasoning: as such CoTs are often synthesized from textual captions or other text-based descriptions rather than directly from the acoustic signal, their alignment with the underlying audio evidence is not explicitly ensured \cite{stepaudior1, audioreasoningsurvey}. This mismatch may contribute to hallucinations. This concern is particularly relevant to audio, where task-critical information may reside in prosody, timbre, overlapping events, temporal structure, background acoustics, and cross-source relationships that cannot always be faithfully reduced to a textual description.

Latent reasoning provides an alternative to this text-centric paradigm by allowing reasoning to take place in a continuous representation space \cite{coconut, flair}. Yet latent reasoning alone does not address the modality-gap issue: without an appropriate acoustic target space, the learned latent reasoning trajectory may still encode predominantly linguistic or short-cut information. Therefore, the key challenge is not simply to move reasoning from text into a continuous space, but to construct latent supervision that incorporates representations derived directly from the acoustic signal, rather than relying solely on textual reasoning traces. However, many audio encoders used in LALMs, such as Whisper \cite{whisper}, are optimized with audio-to-text prediction objectives \cite{liu2025aligning, liu2026code}. Although effective for language-conditioned audio understanding, their representations are shaped toward linguistic prediction and may not provide an ideal target space for supervising acoustically conditioned latent reasoning.

This motivates using an acoustic representation space learned without text-prediction supervision. Joint-Embedding Predictive Architectures (JEPAs) learn representations by predicting latent targets from contextual observations, rather than reconstructing raw inputs or predicting textual labels. WavJEPA \cite{wavjepa} extends this predictive objective to raw waveforms and captures temporally predictable acoustic structure without transcript, caption, or answer supervision. Such predictive waveform representations provide a promising space for constructing latent reasoning trajectories.

Based on this insight, we introduce \textbf{JELAR}, a \textbf{JE}PA-conditioned \textbf{L}atent \textbf{A}udio \textbf{R}easoning framework for LALMs. During training, a frozen WavJEPA encoder extracts audio representations from the input waveform. A non-causal expert uses the question and ground-truth answer to construct a set of queries, which cross-attend to the WavJEPA embeddings to produce fixed-length latent reasoning targets. At inference, the LALM independently generates the fixed-length latent reasoning sequence and subsequently produces the textual answer. Experiments on MMAU-mini and MMAR demonstrate that JELAR improves average accuracy over the Audio-Reasoner baseline by 2.70 and 9.10 absolute percentage points. Our contributions are summarized as follows:
\begin{itemize}
    \item We introduce JELAR, a latent reasoning framework that replaces explicit textual CoT supervision with continuous latent targets conditioned on JEPA-based waveform representations.

    \item We incorporate WavJEPA acoustic representations into a training-only non-causal expert to construct acoustically informed latent targets, which the LALM learns to autoregressively predict without the expert or WavJEPA at inference.

    \item Controlled experiments on MMAU-mini and MMAR show consistent gains over explicit CoT training, while ablations isolate the contributions of acoustic conditioning, WavJEPA-based targets, and latent-only reasoning.
\end{itemize}

\begin{figure*}[t]

\begin{minipage}[b]{1.0\linewidth}
  \centering
  \centerline{\includegraphics[width=15.5cm]{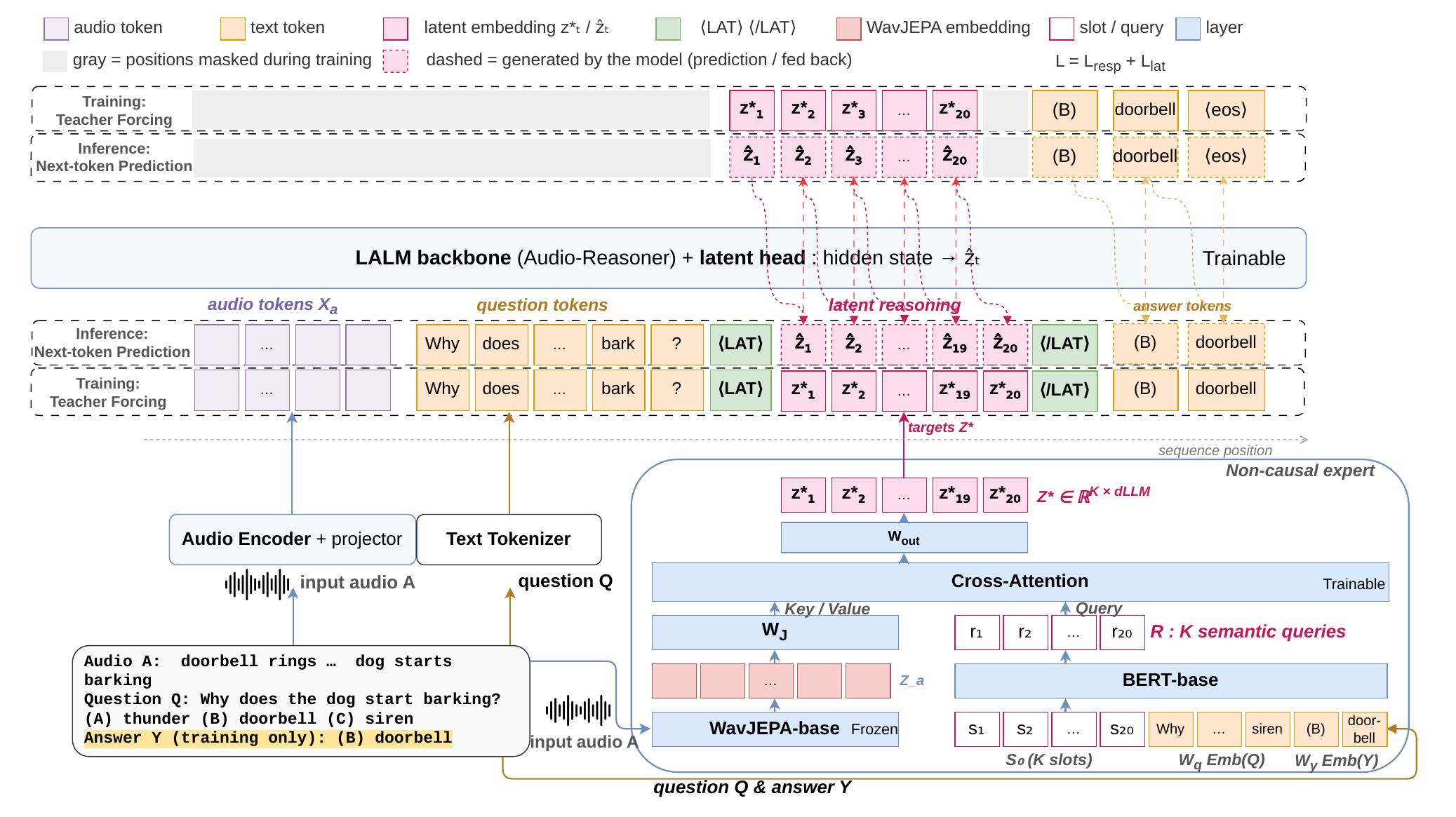}}
\end{minipage}
\caption{The training and inference pipeline of proposed JELAR.}
\label{fig:architecture}
\vspace{-10pt}
\end{figure*}

\section{Method}
\label{sec:method}

\subsection{Overview}
\label{ssec:overview}
The architecture of the proposed JELAR is shown in Figure \ref{fig:architecture}. Given an audio waveform
$a$ and a question $q$, the native audio encoder and the projector of the LALM first convert
the waveform into a sequence of audio tokens:
\begin{equation}
    X_a
    =
    E_{\mathrm{A}}(a),
    \qquad
    X_a\in\mathbb{R}^{N_x\times d_{\mathrm{LLM}}},
\label{eq:native_audio_tokens}
\end{equation}
where $N_x$ is the number of audio tokens and $d_{\mathrm{LLM}}$ is the
input embedding dimension of the language model. 

The LALM receives the audio tokens $X_a$ together with the question tokens and
generates a fixed-length sequence of continuous latent embeddings:
\begin{equation}
    \widehat Z
    =
    \left\{
    \widehat z_1,\ldots,\widehat z_K
    \right\},
    \qquad
    \widehat z_t\in\mathbb{R}^{d_{\mathrm{LLM}}},
\end{equation}
where $K$ denotes the latent reasoning steps. Each predicted latent embedding
is fed back to the language model as the input embedding of the next reasoning
step. During training, the latent inputs for subsequent reasoning steps are provided by a non-causal expert rather than by the model’s previous predictions. At inference, the model recursively feeds back its own predicted latent embeddings.

\subsection{Supervised Fine-Tuning (SFT) for Latent Reasoning}
\label{ssec:training}
Since continuous reasoning trajectories are not available in the
training data, we introduce a non-causal expert, consisting of a bidirectional BERT encoder and a cross-attention module, to construct latent reasoning
targets. For each training example $(a,q,y)$, the expert first forms answer-aware
semantic queries from $(q,y)$, which then cross-attend to waveform representations extracted by WavJEPA \cite{wavjepa}. 

\textbf{JEPA-conditioned latent target construction:}
To obtain the acoustic representations, we use a frozen WavJEPA encoder $E_{\mathrm{J}}$ to convert the waveform into a sequence of embeddings \cite{wavjepa}:
\begin{equation}
    Z_a
    =
    E_{\mathrm{J}}(a),
    \qquad
    Z_a\in\mathbb{R}^{N_a\times d_{\mathrm{J}}},
\label{eq:wavjepa_tokens}
\end{equation}
where $N_a$ and $d_{\mathrm{J}}$ denote the sequence length and feature
dimension of the WavJEPA representations, respectively. To determine which acoustic information is relevant to the target answer,
we first construct $K$ answer-aware semantic queries by a BERT encoder \cite{bert}. Let $S_0\in
\mathbb{R}^{K\times d_{\mathrm{B}}}$ denote $K$ learnable token slots, where
$d_{\mathrm{B}}$ is the hidden dimension of the bidirectional BERT.
The question and answer embeddings are processed jointly with these slots:
\begin{equation}
\begin{split}
    R
    =
    \operatorname{BERT}
    \Big(
    [
    S_0;\,
    W_q(\operatorname{Emb}(q));\,
    W_y(\operatorname{Emb}(y))
    ]
    \Big)_{1:K},
\end{split}
\label{eq:semantic_queries}
\end{equation}
where $W_{\mathrm{q}}$ and $W_{\mathrm{y}}$ project the question and ground-truth answer embeddings into the BERT
dimension and the semantic queries $R=\{r_1,\ldots,r_K\}$ consist of the output states corresponding to the
learnable slots. The semantic queries subsequently attend to the WavJEPA representation sequence through a cross-attention module:
\begin{equation}
    G
    =
    \operatorname{CrossAttn}
    \left(
    Q=R,\,
    K=W_{\mathrm{J}}Z_a,\,
    V=W_{\mathrm{J}}Z_a
    \right),
\label{eq:cross_attention}
\end{equation}
where $W_{\mathrm{J}}$ projects the WavJEPA representations into the query
dimension. The representations are finally projected into the LALM
embedding space:
\begin{equation}
    Z^*
    =
    \left\{
    z^*_1,\ldots,z^*_K
    \right\}
    =
    W_{\mathrm{out}}G, \qquad Z^*\in\mathbb{R}^{K\times d_{\mathrm{LLM}}}.
\label{eq:latent_targets}
\end{equation}

These embeddings constitute the JEPA-conditioned latent targets, combining answer-aware semantic queries with acoustic information retrieved from the predictive waveform representations.

\noindent\textbf{Training objective.}
During teacher-forced training, $Z^*$ is inserted between
$\langle\mathrm{LAT}\rangle$ and $\langle/\mathrm{LAT}\rangle$ tokens. Let $f_{\mathrm{lat}}$
be a latent prediction head. The hidden state at the $\langle\mathrm{LAT}\rangle$ position predicts the
first latent embedding, while the hidden state at the position of
$z^*_{t-1}$ predicts the next one:
\begin{equation}
    \widehat z_t
    =
    f_{\mathrm{lat}}
    \left(
    h_{z^*_{t-1}}
    \right),
    \quad
    t=2,\ldots,K.
\label{eq:latent_prediction}
\end{equation}

We optimize the latent predictions using Smooth L1 loss:
\begin{equation}
    \mathcal{L}_{\mathrm{lat}}
    =
    \frac{1}{K}
    \sum_{t=1}^{K}
    \operatorname{SmoothL1}
    \left(
    \widehat z_t,\,
    \operatorname{sg}(z^*_t)
    \right).
\label{eq:latent_loss}
\end{equation}
Stop-gradient is applied to the latent target in
Eq.~\eqref{eq:latent_loss}.

The response loss is computed over the ground-truth answer tokens:
\begin{equation}
    \mathcal{L}_{\mathrm{resp}}
    =
    -\frac{1}{|y|}
    \sum_{i=1}^{|y|}
    \log
    p_{\theta}
    \left(
    y_i
    \mid
    X_a,q,Z^*,y_{<i}
    \right).
\label{eq:response_loss}
\end{equation}
$Z^*$ is inserted into the LALM without detachment, thus
Eq.~\eqref{eq:response_loss} also optimizes the non-causal expert, encouraging
it to construct latent targets that support response generation.

The complete training objective is
\begin{equation}
    \mathcal{L}
    =
    \mathcal{L}_{\mathrm{resp}}
    +
    \mathcal{L}_{\mathrm{lat}}.
\label{eq:total_loss}
\end{equation}
The WavJEPA encoder remains frozen, while all other components, including the non-causal expert, latent prediction head, audio encoder and projector, and LLM backbone, are jointly optimized.

\subsection{Inference}
\label{ssec:inference}

At inference time, the non-causal expert and WavJEPA encoder are discarded.
Given an audio waveform $a$ and a question $q$, as well as a special token $\langle\mathrm{LAT}\rangle$, the LALM first generates the latent reasoning embeddings recursively. At step
$t$, the final hidden state is mapped to a continuous
latent embedding $\widehat z_t$, 
which is fed back to LLM at the next time step $t+1$. After $K$ latent steps, the
$\langle/\mathrm{LAT}\rangle$ token is appended. The LALM then continues
standard autoregressive decoding to generate the formatted textual response:
\begin{equation}
    \widehat y_i
    \sim
    p_{\theta}
    \left(
    \cdot
    \mid
    X_a,q,\widehat Z,\widehat y_{<i}
    \right).
\end{equation}

\begin{table*}[t]
\centering
\caption{Accuracy (\%) on MMAR. Results marked with $\dagger$ are taken from
the MMAR benchmark paper~\cite{mmar}; S, M, and Sp denote sound, music, and speech, respectively. Boldface denotes the better-performing result only within each controlled comparison.}
\label{tab:mmar_main}
\renewcommand{\arraystretch}{0.8}
\begin{tabular}{lcccccccc}
\toprule
Model &
S &
M &
Sp &
S--M &
S--Sp &
M--Sp &
S--M--Sp &
Avg. \\
\midrule
Qwen2-Audio-Instruct$^\dagger$
& 33.33 & 24.27 & 32.31 & 9.09
& 31.19 & 30.49 & 25.00 & 30.00 \\

Qwen-2.5-Omni-7B$^\dagger$
& 58.79 & 40.78 & 59.86 & 54.55
& 61.93 & 67.07 & 58.33 & 56.70 \\

GPT-4o Audio$^\dagger$
& 53.94 & 50.97 & 70.41 & 63.64
& 72.48 & 62.20 & 75.00 & 63.50 \\

Gemini 2.0 Flash$^\dagger$
& 61.21 & 50.97 & 72.11 & 81.82
& 72.48 & 65.85 & 70.83 & 65.60 \\
\midrule
Audio-Reasoner
& 44.24 & 35.92 & 46.26 & 27.27
& 47.71 & 50.00 & 25.00 & 43.70 \\

\textbf{JELAR}
& \textbf{53.33} & \textbf{49.51} & \textbf{51.70} & \textbf{45.45}
& \textbf{57.34} & \textbf{53.66} & \textbf{50.00} & \textbf{52.80} \\
\bottomrule
\end{tabular}
\vspace{-10pt}
\end{table*}

\section{Experiments}
\label{sec:experiments}

\subsection{Experimental Setup}
\label{ssec:experimental_setup}

\noindent\textbf{Model configuration.}
We build JELAR on the same Audio-Reasoner backbone as the baseline \cite{audioreasoner}. The explicit-CoT baseline and JELAR use the same training data and optimization configuration. The key difference is the reasoning supervision: Audio-Reasoner is trained with explicit textual CoT, whereas JELAR replaces textual CoT with continuous latent reasoning. We therefore emphasize their controlled comparison to assess the effect of replacing explicit CoT with JEPA-conditioned latent reasoning. In JELAR, a frozen WavJEPA-base encoder\footnote{\url{https://huggingface.co/labhamlet/wavjepa-base}}
provides the representations used to construct the latent targets, while the answer-aware query constructor is initialized from BERT-base-uncased\footnote{\url{https://huggingface.co/google-bert/bert-base-uncased}}.
Unless otherwise specified, the latent reasoning stage contains $K=20$ embeddings. We choose this based on performance on a held-out validation split of the development data.

We evaluate the models on MMAU-mini \cite{mmau} and
MMAR \cite{mmar}. MMAU-mini consists of closed-choice questions covering sound, music, and speech. MMAR contains human-curated multiple-choice questions designed for \textbf{multi-step audio reasoning}.
\vspace{-2pt}
\subsection{Main Results}
\label{ssec:main_results}

\begin{table}[t]
\centering
\caption{Accuracy (\%) on MMAU-mini. Results marked with $\dagger$ are
reported in \cite{audioreasoner}. The bottom two rows constitute the primary controlled
comparison.}
\label{tab:mmau_main}
\renewcommand{\arraystretch}{0.85}
\resizebox{\columnwidth}{!}{
\begin{tabular}{lcccc}
\toprule
Model & Sound & Music & Speech & Avg. \\
\midrule
GPT-4o + Caption$^\dagger$
    & 63.36 & 60.77 & 53.15 & 57.30 \\
Gemini-1.5-Pro$^\dagger$
    & 56.75 & 49.40 & 58.55 & 54.90 \\
Qwen2-Audio-Instruct$^\dagger$
    & 54.95 & 50.98 & 42.04 & 49.20 \\
\midrule
Audio-Reasoner
    & \textbf{66.86} & 61.57 & 51.37 & 59.70 \\
\textbf{JELAR}
    & 64.31 & \textbf{62.28} & \textbf{60.66} & \textbf{62.40} \\
\bottomrule
\end{tabular}}
\vspace{-10pt}
\end{table}

\noindent\textbf{MMAU-mini results.}
As shown in Table \ref{tab:mmau_main}, JELAR improves the average accuracy of Audio-Reasoner from 59.70\% to 62.40\%. The gain is mainly driven by the speech domain, which increases from 51.37\% to 60.66\%, while music improves slightly and sound accuracy decreases.

\noindent\textbf{MMAR results.}
Table~\ref{tab:mmar_main} shows that JELAR delivers a substantially larger gain on MMAR, raising average accuracy from 43.70\% to 52.80\%, an absolute improvement of 9.10 percentage points. Notably, it outperforms Audio-Reasoner across all seven categories, with especially large gains on mixed-domain reasoning tasks. 
The gain on MMAR is considerably larger than that on MMAU-mini. This is
consistent with the design of MMAR, which places greater emphasis on multi-step
reasoning and mixed-domain audio reasoning. The larger gain on MMAR is also consistent with the hypothesis that the proposed latent reasoning may be particularly beneficial for tasks requiring integration of multiple acoustic cues.

\begin{table}[t]
\centering
\caption{Effect of the latent reasoning length $K$. We report average
classification accuracy (\%) on MMAU-mini and MMAR.}
\label{tab:latent_length}
\resizebox{\columnwidth}{!}{
\begin{tabular}{lccccccc}
\toprule
$K$ & 8 & 12 & 16 & \textbf{20} & 24 & 28 & 32 \\
\midrule
MMAU-mini
& 60.60 & 61.00 & 61.80 & \textbf{62.40}
& 60.00 & 61.10 & 60.80 \\

MMAR
& 49.60 & 50.40 & 51.00 & \textbf{52.80}
& 51.00 & 51.30 & 51.70 \\
\bottomrule
\end{tabular}}
\vspace{-10pt}
\end{table}
\begin{table}[t]
\centering
\caption{Ablation of acoustic conditioning and reasoning format.}
\label{tab:ablation}
\renewcommand{\arraystretch}{0.9}
\begin{tabular}{lcc}
\toprule
Variant & MMAU & MMAR \\
\midrule
\textbf{JELAR (WavJEPA, latent-only)} & \textbf{62.40} & \textbf{52.80} \\
w/o acoustic conditioning, latent-only & 60.50 &47.60 \\
Whisper encoder, latent-only & 61.60 & 50.70 \\
WavJEPA, latent + explicit CoT & 61.40 & 49.30 \\
\bottomrule
\end{tabular}
\vspace{-10pt}
\end{table}
\vspace{-5pt}
\subsection{Ablation study}
\label{ssec:latent_length}

Table \ref{tab:latent_length} examines the effect of latent reasoning length. Increasing \(K\) from 8 to 20 improves MMAU-mini from 60.60\% to 62.40\% and MMAR from 49.60\% to 52.80\%, with both benchmarks peaking at \(K=20\). Increasing \(K\) further does not yield additional gains, suggesting that more latent reasoning steps are not inherently beneficial once sufficient reasoning capacity is reached.

Table \ref{tab:ablation} analyses the contributions of acoustic conditioning and reasoning format. Removing acoustic conditioning reduces the accuracy to 60.50\% on MMAU-mini and 47.60\% on MMAR. Conditioning the latent targets on the native Whisper encoder of Audio-Reasoner \cite{whisper, audioreasoner} improves the results to 61.60\% and 50.70\%, while WavJEPA further raises them to 62.40\% and 52.80\%. These results indicate that acoustic conditioning provides additional benefit beyond answer-aware latent supervision, while WavJEPA provides a more effective acoustic target representation than the native Whisper encoder in this setting.

For reasoning format, explicit textual CoT provides no additional benefit once latent reasoning is introduced. Adding an explicit CoT stage after the latent sequence lowers performance from 62.40\% to 61.40\% on MMAU-mini and from 52.80\% to 49.30\% on MMAR. This supports latent-only reasoning as the more effective reasoning format in our setting.
\section{Conclusion}
\label{sec:conclusion}

In this work, we introduced JELAR, a JEPA-conditioned latent reasoning framework for LALMs. JELAR replaces explicit textual CoT supervision with continuous latent targets constructed by answer-aware semantic queries attending to frozen WavJEPA representations. Under controlled comparisons with Audio-Reasoner, JELAR improves average accuracy by 2.70 and 9.10 percentage points on MMAU-mini and MMAR, respectively. Ablation studies further support the contributions of acoustic conditioning, WavJEPA-based target representations, and latent-only reasoning. Future work will investigate larger backbones and more diverse audio reasoning data.

\vfill\pagebreak

\bibliographystyle{IEEEbib}
\bibliography{strings,refs}

\end{document}